\documentclass[epj]{svjour}
\usepackage{graphicx,mathrsfs,amsmath,bm,amssymb}
\usepackage{physics}
\usepackage{hyperref}
\usepackage{multirow}
\usepackage{lscape}
\usepackage{color}
\usepackage{hyperref}
\usepackage{slashed}
\begin{document}

\title{A new class of hybrid EoS with multiple critical endpoints for simulations of supernovae, neutron stars and their mergers}

\author{
O. Ivanytskyi\inst{1}
\and
D. Blaschke\inst{1}
}

\institute{
Institute of Theoretical Physics, University of Wroclaw, 
Max Born place 9, 50-204 Wroclaw, Poland 
}
\date{Received: date / Revised version: date}
%
\abstract{
We introduce a family of equations of state (EoS) for hybrid neutron star (NS) matter that is obtained by a two-zone parabolic interpolation between a soft hadronic EoS at low densities and a stiff quark matter EoS with color superconductivity at high densities within a finite region of baryonic chemical potentials $\mu_B^h < \mu_B < \mu_B^q$. 
We consider two scenarios corresponding to a cross-over and a strong 
first-order transition between quark and hadron phases considered at finite and zero temperatures. This allows us to analyze the effects of finite entropy on the EoS and mass-radius relation of NS. 
We demonstrate that the formation of a color superconducting state of quark matter drives the evolution of matter in supernovae explosions under the condition of entropy conservation to higher temperatures than in the case of deconfinement to normal quark matter.
Within the presented hybrid EoS scenario, regions of the QCD phase diagram may be accessible to supernovae and NS mergers that can be reached also in terrestrial experiments with relativistic heavy ion collisions.
} 
\maketitle
\section{Introduction}
\label{sec:intro}
Simulations of core-collapse supernova (SN) explosions and binary neutron star (BNS) mergers with model equations of state (EoS) are a unique tool to investigate the QCD phase diagram in the region of low temperatures and high baryon densities
($T\lesssim 60$ MeV at $1\lesssim n/n_0 \lesssim 5$) 
which is otherwise inaccessible, in particular to lattice QCD simulations and heavy-ion collision (HIC) experiments
\cite{Bauswein:2022vtq}.
Not only that the detection of signals of a strong first-order phase transition in BNS mergers \cite{Bauswein:2018bma,Bauswein:2020aag}
and/or supernovae \cite{Fischer:2017lag,Fischer:2021tvv}
would provide support for the existence of a CEP in the phase diagram that has so far been unsuccessfully been sought for in HIC experiments, there is theoretical evidence for a crossover transition at very low temperatures that suggests the existence of a second CEP
or even a crossover-all-over situation.  
This arises from the observation \cite{Hatsuda:2006ps,Abuki:2010jq} that a coexistence of chiral symmetry breaking and diquark condensation occurs at low temperatures due to the $U_A(1)$ 
anomaly-generating triangle diagram which, after Fierz transformation mixes diquark and meson condensates. 
This effect realizes the concept of quark-hadron continuity \cite{Schafer:1998ef} in a crossover transition. 
Support for such a picture comes also from recent progress in NS phenomenology.
\\
We are currently witnessing a paradigm change in the interpretation of mass and radius measurements of pulsars that is induced by the observation that from the multi-messenger analysis of typical-mass neutron star radii with $R_{1.4~M_\odot}=11.7^{+0.86}_{-0.81}$ km 
\cite{Dietrich:2020efo} (see also \cite{Capano:2019eae}) and the recent NICER radius measurement $R_{2.0~M_\odot}=13.7^{+2.6}_{-1.5}$ km \cite{Miller:2021qha} (see also \cite{Riley:2021pdl}) follows that $R_{2.0~M_\odot}\gtrsim R_{1.4~M_\odot}$.
The description of such a behaviour as solution of the Tolman-Oppenheimer-Volkoff (TOV) equations requires a soft-stiff transition in the EOS at densities $n\lesssim 2 n_0$, just before the hyperon onset. 
This transition could be the hadron-to-quark matter transition.
In recent descriptions one joins a standard nuclear EOS with a 
constant speed of sound (CSS) model for the high-density phase either by a first-order phase transition (with a vanishing speed of sound
in the mixed phase \cite{Somasundaram:2021ljr}) or by directly matching the nuclear and quark matter squared speed of sound $c_s^2$ at a certain transition density $n_{\rm tr}$ without a density jump, thus mimicking a crossover transition \cite{Drischler:2020fvz}.
The best phenomenological description fulfilling simultaneously 
the constraints on both radii $R_{2.0~M_\odot}$
and $R_{1.4~M_\odot}$ is obtained in this simple picture by 
$n_{\rm tr}\sim 0.5~n_0$ and $c_s^2\sim 0.5$. 
We would like to remark that a CSS model with 
$c_s^2= 0.45 \dots 0.54$ provides an excellent fit to a microscopic nonlocal chiral quark model of the Nambu--Jona-Lasinio (NJL) type 
with diquark condensation (color superconductivity) and repulsive vector meson mean field \cite{Antic:2021zbn,Contrera:2022tqh}.
A direct, one-zone interpolation scheme between the safely known soft nuclear matter EoS (up to about $1.1~n_0$ as in \cite{Hebeler:2013nza})
and the suitably chosen stiff quark matter EoS (e.g., from a NJL model with coupling to a repulsive vector meson mean field) was pioneered in the works of \cite{Masuda:2012kf,Masuda:2012ed}.
Such a phase transition construction can be understood as a shortcut
for three physical effects as ingredients: 
\begin{enumerate}
\item[(i)] a stiffening of the nuclear matter EoS $P_H(\mu)$ due to the repulsive quark Pauli blocking effect between nucleons \cite{Ropke:1986qs,Blaschke:2020qrs} which can be effectively accounted for with a nucleonic excluded volume (see, e.g., \cite{Alvarez-Castillo:2016oln}), 
\item[(ii)] a strong reduction of the quark matter pressure $P_Q(\mu)$ at low chemical potentials due to confining forces (which result then in a good crossing of curves at $\mu=\mu_c$ allowing for the Maxwell construction $P_H(\mu_c)=P_Q(\mu_c)$ of a first-order phase transition), and
\item[(iii)] a mixed phase construction (e.g., by a parabolic interpolation   \cite{Ayriyan:2017nby}) that mimics the effects of finite-size structures (pasta phases) in the quark-hadron coexistence region. 
\end{enumerate}
For more details on the physics background, see \cite{Ayriyan:2021prr,Baym:2017whm} and and references therein.
With this microphysical basis behind the interpolation approach, a two-zone interpolation scheme (TZIS) for the hadron-to-quark matter has been developed in \cite{Ayriyan:2021prr}, where at the matching point $\mu_c$ situated between $\mu_H$ and $\mu_Q$, one can choose the condition of continuous density ($\Delta n=0$, crossover) or a finite density jump ($\Delta n \neq 0$, first-order transition). 
In the present work, we present a generalization of this TZIS to finite temperatures (and arbitrary isospin densities) as a necessary prerequisite for investigating the consequences of these recent developments in the interpretation of neutron star phenomenology at zero temperature to simulations of supernova explosions and of binary neutron star merger events. 
The goal is to model the general class of hybrid EoS that corresponds to a phase diagram which has not only one critical endpoint (CEP) at high temperatures which marks the change from a first-order to a crossover transition regime, but also second CEP at low temperatures that arises from the competition and mixing between dynamical chiral symmetry breaking and color superconductivity.
Within the finite-temperature generalization of the TZIS, this can be achieved by defining the function 
$\Delta n[\mu_c(T)]$ along the matching line $\mu_c(T)$ between the hadron-like and the quark-like interpolation zone in the phase diagram.  
The function $\Delta n[\mu_c(T)]$ encodes the position of the CEPs
$T_{\rm cep1}$ and $T_{\rm cep2}$ where $\Delta n=0$ as well as the strength of the first-order transition between these points where  
$\Delta n \neq 0$.
\\
It is the aim of our ongoing research to investigate the dependence of the above described signals of a strong phase transition in supernova explosions and binary neutron star mergers on the detailed structure of the QCD phase diagram at low temperatures and high baryon densities and thus to be prepared for interpreting the possible observation of signals from such events in the near future.
\section{Quark matter equation of state}
\label{sec2}

Here we outline the main aspects of the quark matter EoS. The interested readers are addressed to Refs. \cite{Ivanytskyi:2021dgq,Ivanytskyi:2022oxv} where the model was developed. Its is a chirally symmetric formulation of the density functional approach to quark matter \cite{Kaltenborn:2017hus}, which allows scalar diquark pairing leading to the phenomenon of color superconductivity. In the two flavor case considered here such pairing leads to formation of the 2SC phase of quark matter. Note, the three flavor case leading to formation of the color-flavor locked (CFL) quark matter was considered within the present approach in Ref. \cite{Blaschke:2022knl}. The model is represented by the Lagrangian
\begin{eqnarray}
\label{I}
\mathcal{L}&=&\overline{q}(i\slashed\partial- m)q+\mathcal{L}_V+\mathcal{L}_D-\mathcal{U}.
\end{eqnarray}
Quark fields are described by the flavor spinor $q^T=(u,d)$ and $m$ is the current mass. Vector repulsion and diquark paring interactions enter Eq. (\ref{I}) trough
\begin{eqnarray}
\label{II}
\mathcal{L}_V&=&-G_V(\overline{q}\gamma_\mu q)^2,\\
\label{III}
\mathcal{L}_D&=&G_D
(\overline{q}i\gamma_5\tau_2\lambda_A q^c)(\overline{q}^ci\gamma_5\tau_2\lambda_A q)
\end{eqnarray}
with $G_V$ and $G_D$ being coupling constants. Attractive interaction in scalar and pseudoscalar channels is given by the potential 
\begin{eqnarray}
\label{IV}
\mathcal{U}=D_0\left[(1+\alpha)\langle \overline{q}q\rangle_0^2
-(\overline{q}q)^2-(\overline{q}i\gamma_5\vec\tau q)^2\right]^{\frac{1}{3}},
\end{eqnarray}
where $\langle \overline{q}q\rangle_0$ is vacuum value of chiral condensate, while constant $D_0$ and $\alpha$ control the interaction strength and constituent quark mass in the vacuum \cite{Ivanytskyi:2021dgq,Ivanytskyi:2022oxv}, respectively. This potential respects chiral symmetry of strong interaction. It can be expanded around the mean-field solutions $\langle \overline{q}q\rangle$ and $\langle \overline{q}i\gamma_5\vec\tau q\rangle=0$. In what follows the subscript indes ``$MF$'' labels the quantities defined at the mean field. The second order expansion of $\mathcal{U}$ implies the following non-vanishing expansion coefficients 
\begin{eqnarray}
\label{V}
\Sigma_{MF}&=&\frac{\partial\mathcal{U}_{MF}}{\partial\langle\overline{q}q\rangle},\\
\label{VI}
G_S&=&-\frac{1}{2}
\frac{\partial^2\mathcal{U}_{MF}}{\partial\langle\overline{q}q\rangle^2},\\
\label{VII}
G_{PS}&=&-\frac{1}{6}
\frac{\partial^2\mathcal{U}_{MF}}{\partial\langle\overline{q}i\gamma_5\vec\tau q\rangle^2}.
\end{eqnarray}
This brings the Lagrangian to the effective current-current interaction form of the NJL model type
\begin{eqnarray}
\mathcal{L}^{eff}&=&\overline{q}(i\slashed\partial- m^*)q+G_S(\overline{q}q-\langle\overline{q}q\rangle)^2+G_{PS}(\overline{q}i\gamma_5\vec\tau q)^2\nonumber\\
\label{VIII}
&+&\mathcal{L}_V+\mathcal{L}_D-\mathcal{U}_{MF}+\langle\overline{q}q\rangle\Sigma_{MF}.
\end{eqnarray}
Here $m^*=m+\Sigma_{MF}$ is the constituent quark mass. Its form allows us to interpret $\Sigma_{MF}$ as a mean-field self-energy of quarks. On the other hand, it follows from the form of the scalar and pseudoscalar interaction channels in Eq. (\ref{VIII}) that $G_{S}$ and $G_{PS}$ are nothing else as the corresponding effective couplings. They are medium dependent and differ in the general case. This signals about violation of chiral symmetry. This violation is a direct sequence of expanding $\mathcal{U}$ around the mean-field solution,which is know be chirally broken. However, at high densities and temperatures $G_{S}$ and $G_{PS}$ asymptotically coincide being a consequence of the dynamical restoration of chiral symmetry \cite{Ivanytskyi:2021dgq,Ivanytskyi:2022oxv}.

In Ref. \cite{Ivanytskyi:2022oxv} parameters of the present model were fixed using the strategy typical for chiral models of quark matter, i.e. by fitting them to vacuum values of the quantities relevant to QCD phenomenology. The most important of them are mass $M_\pi$ and decay constant $F_\pi$ of the pseudoscalar mode representing pion. The scalar mode mass $M_\sigma$ also was considered in the respect. However, the experimental status of the corresponding meson is far from being clear. Therefore, $M_\sigma$ was allowed to vary around the mass of $f_0(980)$ meson. Note, the lightest candidate for the scalar meson role $f_0(500)$ was not considered due to its high width about 500-1000 MeV \cite{PhysRevD.98.030001}. Our approach as well as the most of chiral models of quark matter \cite{Grigorian:2006qe} is unable to reproduce the vacuum value of chiral condensate per flavor $|\langle\overline{l}l\rangle^{1~GeV}_0|^{1/3}=241$ MeV found from QCD sum rules at the renormalization scale 1 GeV \cite{Jamin:2002ev}. In order to fix a compromised value of this quantity it was analyzed together with the pseudocritical temperature $T_{PC}$ defined by the peak position of chiral susceptibility.
In addition to the current quark mass $m$ and interaction potential parameters $D_0$ and $\alpha$ the present model includes momentum scale $\Lambda$, which regularizes zero point terms in the expression for the thermodynamic potential (see Ref. \cite{Ivanytskyi:2022oxv} for details). Table \ref{table1} shows values of these parameters, which reproduce $M_\pi=140$ MeV, $F_\pi=90$ MeV, $M_\sigma=980$ MeV, $|\langle\overline{l}l\rangle^{1~GeV}_0|=267$ MeV and $T_{PC}=163$ MeV. This parameterization of the present model yields $m^*=718$ MeV in the vacuum. Such high value of the constituent quark mass provides an efficient phenomenological confinement of quarks at low temperatures and densities.

The values of vector $G_V$ and diquark $G_D$ pairing constants from Table \ref{table1} were adjusted in order to provide the best agreement with the observational constraints on the mass-radius diagram of compact stars with quark cores \cite{Antoniadis:2013pzd,Riley:2021pdl,Miller:2021qha,Riley:2019yda,Raaijmakers:2019qny,LIGOScientific:2018cki,Bauswein:2017vtn,Annala:2017llu} and their tidal deformabilities \cite{LIGOScientific:2018cki}. 

For the chosen parameter set EoS of quark matter is obtained by applying the mean-field approximation to the effective Lagrangian (\ref{VIII}). It is remarkable that within the density range from two to ten normal nuclear densities variation of squared speed of sound of the present model $c_S^2=0.57 - 0.60$ is just 5 \%. This surves as a microscopic justification of the CSS parametrization of the quark matter EoS.

\begin{table}[t]
\begin{tabular}{|c|c|c|c|c|c|c|c|c|c|c|c|}
\hline
  $m$ [MeV] & $\Lambda$ [MeV] & $\alpha$ & $D_0\Lambda^{-2}$ &
  $G_V\Lambda^2$ & $G_D\Lambda^2$ \\ \hline 
      4.2   &    573    &    1.43  &  1.39  & 1.58 & 3.30  \\ \hline 
\end{tabular}
\caption{Parameters of the present model of quark matter EoS.}
\label{table1}
\end{table}
%

\section{Quark-hadron transition}
\label{sec3}

Quark degrees of freedom are relevant to description of strongly interacting matter only at high densities, while in the low density regime they are confined and hadronized. This requires description of strongly interaction matter in the mentioned regime with a hadronic EoS. For this we use the DD2 EoS \cite{Typel:2009sy}. Hybrid quark-hadron EoS is obtained by merging the one phase quark and hadron EoS according to a given construction of phase transition. In this work we consider two constructions of quark-hadron transition described below.

These constructions require pressures of hadron and quark pressures as functions of baryonic $\mu_B$ and electric $\mu_Q$ chemical potentials. Note, the strange chemical potential does not appear since we consider the two flavor case. Furthermore, requiring a given value of the electric charge fraction $Y_Q=\frac{n_Q}{n_B}$ baryonic chemical potential becomes the only independent quantity, while electric chemical potential becomes a function of it, i.e. $\mu_Q=\mu_Q(\mu_B)$. The temperature dependence is omitted below for shortening the notations

\subsection{Maxwell construction}

Gibbs criterion of phase equilibrium implies equality of pressures, temperatures and two chemical potentials of quark and hadron phases, which defines the phase coexistence surface.
The Maxwell construction of phase transition between hadrons (superscript index ``$h$'') and quarks (superscript index ``$q$'') relaxes the Gibbs criterion by requiring equality of only baryonic chemical potential (see the recent review \cite{Baym:2017whm}), i.e. the $\mu_B^h=\mu_B^q\equiv\mu_B^{max}$. Hereaftre the subscript index ``$max$'' denotes the quantities defined at this value of the baryonic chemical potential. Thus, the criterion of phase equilibrium becomes 
\begin{eqnarray}
\label{IX}
p^h|_{max}=p^q|_{max}.
\end{eqnarray}
The characteristic feature of the Maxwell construction is a discontinuous density jump signalling about strong first order phase transition. Indeed, defining a given charge density as a partial derivative of pressure with respect to the corresponding chemical potential we immediately conclude that in the general case
\begin{eqnarray}
\label{X}
n^h_{B,Q}|_{max}\neq n^q_{B,Q}|_{max}.
\end{eqnarray}
This discontinuous change of density is caused by a sharp interface between quark and hadron phases due to high surface tension leading to separation between them. It leads to a flat plateau like shape of the mixed phase in the density pressure-plane. Electric chemical potential entering this relation also experiences a discontinuous jump at the transition between two phases
\begin{eqnarray}
\label{XI}
\mu_Q^h|_{max}\neq \mu_Q^q|_{max}.
\end{eqnarray}
%

\subsection{Two-zone interpolation scheme}

Discontinuity of electric chemical potential is a well known pitfall of the Maxwell construction. It can be removed by an accurate incorporation of the full Gibbs criterion also known as the Glendenning construction \cite{Glendenning:1992vb}. As a results flat shape of the mixed phase region gets washed out, while $\mu_Q$, $n_B$ and $n_Q$ become continuous functions of $\mu_B$. In this case the mixed phase itself is a homogeneous mixture of the quark and hadron ones, which is possible only at vanishing surface tension of their interface. In a realistic case surface tension lays between vanishing Glendenning and high Maxwell values. Interplay between its effects and Coulomb interaction leads to formation of inhomogeneous finite size structures known as pastas \cite{Maslov:2018ghi}. Replacement interpolation construction provides an efficient and simple way to mimic inhomogeneous mixed phase of quarks and hadrons \cite{Ayriyan:2017nby}. It, however, does not allow a strong first order phase transition accompanied by discontinuous density jump. In order to consider such possibility in this work we use TZIS \cite{Ayriyan:2021prr}, which also effectively accounts for the effects of stiffening of hadron EoS due to Pauli blocking. Technically, this method corresponds to merging under certain conditions discussed below two parabolic interpolating functions. In Ref. \cite{Ayriyan:2021prr} TZIS was developed for the case of zero change fraction. Here we make the next step and generalize it to finite $Y_Q$, when the mixed phase pressure $\tilde p$ is a function of baryonic $\mu_B$ and electric chemical potentials with $\mu_Q=\tilde\mu_Q(\mu_B)$. Hereafter tilde labels the quantities related to the mixed phase region defined within the TZIS. In order to model a jump of the baryonic density the mixed phase pressure is defined in a peace-wise way as
\begin{eqnarray}
\label{XII}
\tilde p=\left\{
\begin{array}{l}
\tilde p^h\left[\mu_B,\tilde\mu_Q(\mu_B)\right],\quad\mu_B^h\le\mu_B\le \mu_B^c\\
\tilde p^q\left[\mu_B,\tilde\mu_Q(\mu_B)\right],\quad \mu_B^c\le\mu_B\le\mu_B^q
\end{array}
\right..
\end{eqnarray}
where $\mu_B^h$ and $\mu_B^q$ define the edges of the mixed phase interval and for the sake of simplicity merging point is defined symmetrically, i.e.
\begin{eqnarray}
\label{XIII}
\mu_B^c=\frac{\mu_B^h+\mu_B^q}{2}.
\end{eqnarray}
The mixed phase boundary from the hadron side side is parameterized with two constant parameters $x$ and $T_0$ as
\begin{eqnarray}
\label{XIV}
\mu_B^h=\mu_B^{max}|_{T=0}(1-x)\sqrt{1-\frac{T^2}{T_0^2}}
\end{eqnarray}
It is seen that $T_0$ is temperature of the mixed phase onset at zero chemical potential. At this regime quark-hadron transition is a smooth cross-over governed by the restoration of chiral symmetry. The corresponding pseudocritical temperature found in lattice QCD by the position of the peak of chiral susceptibility $C_0^\chi(T)$ is $156.5\pm1.5$ MeV \cite{HotQCD:2018pds}. We estimate $T_0$ to have the value corresponding to the half-heights of the $C_0^\chi$ peak, i.e. $T_0=140$ MeV. On the quark side we parameterize the mixed phase boundary as
\begin{eqnarray}
\label{XV}
\mu_B^{q}=\mu_B^{max}(1+x).
\end{eqnarray}
Eqs. (\ref{XIV}) and (\ref{XV}) provide $\mu_B^c=\mu_B^{max}$ at zero temperature. In this work we consider $x=0.01$.

Partial derivatives of the interpolating pressure with respect to baryon and electric chemical potentials give the densities of the corresponding charges. Using this densities we formally expand quark and hadron branches of $\tilde p$ around the corresponding edges of the mixed phase interval up to the second order. Below we give an explicit treatment to the hadron branch, while the expressions for the quark one can be obtained by replacing all indexes ``$h$'' by ``$q$''. Thus,
\begin{eqnarray}
\tilde p^h\simeq\tilde p^h|_h&+&(\mu_B-\mu_B^h)\left(\tilde n_B^h+\tilde n_Q^h\frac{d\tilde\mu_Q}{d\mu_B}\right)_h
\nonumber\\
&+&\frac{(\mu_B-\mu_B^h)^2}{2}
\left(\frac{\partial\tilde n_B^h}{\partial\mu_B}+
2\frac{\partial\tilde n_Q^h}{\partial\mu_B}\frac{d\tilde\mu_Q}{d\mu_B}
\right.\nonumber\\
\label{XVI}
&+&\left.\frac{\partial\tilde n_Q^h}{\partial\tilde\mu_Q}\left(\frac{d\tilde\mu_Q}{d\mu_B}\right)^2+
\tilde n_Q^h\frac{d^2\tilde\mu_Q}{d\mu_B^2}\right)_h,
\end{eqnarray}   
Hereafter the subscript index ``$h$'' labels the quantities defined at $\mu_B=\mu_B^h$. Similarly, the subscript indexes ``$q$'' and ``$c$'' correspond to the quantities defined at $\mu_B=\mu_B^q$ and $\mu_B=\mu_B^c$, respectively. The densities of baryonic and electric charge densities can be expanded linearly, i.e. 
\begin{eqnarray}
\label{XVII}
\tilde n_B^h
&\simeq&\tilde n_B^h|_h+(\mu_B-\mu_B^h)
\left(\frac{\partial \tilde n_B^h}{\partial\mu_B}+
\frac{\partial\tilde n_B^h}{\partial\tilde\mu_Q}\frac{d\tilde\mu_Q}{d\mu_B}\right)_h,\\
\label{XVIII}
\tilde n_Q^h
&\simeq&\tilde n_Q^h|_h+(\mu_B-\mu_B^h)
\left(\frac{\partial \tilde n_Q^h}{\partial\mu_B}+
\frac{\partial\tilde n_Q^h}{\partial\tilde\mu_Q}\frac{d\tilde\mu_Q}{d\mu_B}\right)_h.
\end{eqnarray}   
Ratio of the densities of electric and baryonic charges yields electric charge fraction $Y_Q=\tilde n_Q^h/\tilde n_B^h$. It can be split into the contributions of baryons $Y_Q^b$ and leptons $Y_Q^l$. In the electrically neutral case $Y_Q=0$ provided by $Y_Q^b=-Y_Q^l$, while for symmetric matter $Y_Q^b=1/2$ and $Y_Q^l=0$. In the case of finite and constant $Y_Q$ one has to require
\begin{eqnarray}
\label{XIX}
\tilde n_Q^h|_h&=&Y_Q\tilde n_B^h|_h,\\
\label{XX}
\left(\frac{\partial \tilde n_Q^h}{\partial\mu_B}+
\frac{\partial\tilde n_Q^h}{\partial\tilde\mu_Q}\frac{d\tilde\mu_Q}{d\mu_B}\right)_h
&=&Y_Q\left(\frac{\partial \tilde n_B^h}{\partial\mu_B}+
\frac{\partial\tilde n_B^h}{\partial\tilde\mu_Q}\frac{d\tilde\mu_Q}{d\mu_B}\right)_h.
\end{eqnarray}   
These relations allow us to parameterize the hadron branch of the interpolating pressure through three independent parameters
\begin{eqnarray}
\label{XXI}
c^h_0&\equiv&\tilde p^h|_h,\\
\label{XXII}
c^h_1&\equiv&\tilde n_B^h|_h,\\
\label{XXIII}
c^h_2&\equiv&\frac{1}{2}\left(\frac{\partial \tilde n_B^h}{\partial\mu_B}+
\frac{\partial\tilde n_B^h}{\partial\tilde\mu_Q}\frac{d\tilde\mu_Q}{d\mu_B}\right)_h.
\end{eqnarray}
As is seen form Eq. (\ref{XVI}), at finite charge fraction this parameterization also requires an information about the dependence of $\tilde\mu_Q$ on $\mu_B$ encoded to the corresponding first and second derivatives evaluated at the mixed phase boundary. This is a new element of the present paper compared to Ref. \cite{Ayriyan:2021prr}. However, the dependence $\tilde\mu_Q=\tilde\mu_Q(\mu_B)$ is not specified by TZIS and should be defined independently. For the sake of simplicity we assume it to be linear, i.e.
\begin{eqnarray}
\label{XXIV}
\tilde\mu_Q=\mu_Q^h+\left(\mu_B-\mu_B^h\right)
\frac{\mu_Q^q|_q-\mu_Q^h|_h}{\mu_B^q-\mu_B^h}.
\end{eqnarray}
This leads to $\frac{d\tilde\mu_Q}{d\mu_B}=const$, $\frac{d\tilde\mu_Q^2}{d\mu_B^2}=0$ and yields
\begin{eqnarray}
\tilde p^{h,q}&=&c^{h,q}_0+(\mu_B-\mu_B^{h,q})\left(1+Y_Q\frac{d\tilde\mu_Q}{d\mu_B}\right)c^{h,q}_1
\nonumber\\
\label{XXV}
& &\hspace*{.6cm}+(\mu_B-\mu_B^{h,q})^2\left(1+Y_Q\frac{d\tilde\mu_Q}{d\mu_B}\right)c^{h,q}_2,\\
\label{XXVI}
\tilde n^{h,q}_B&=&c^{h,q}_1+2(\mu_B-\mu_B^{h,q})c^{h,q}_2.
\end{eqnarray}   
We might naively think that constant factor $1+Y_Q\frac{d\tilde\mu_Q}{d\mu_B}$ in the expression for $\tilde p^{h,q}$ can be absorbed to the expansion coefficients $c_1^{h,q}$ and $c_2^{h,q}$, which makes the present parameterization of the mixed phase pressure identical to the one from Ref. \cite{Ayriyan:2021prr}.
However, this factor is absent in the expression for $\tilde n_B^{h,q}$. Therefore, finite $Y_Q$ can not be excluded from the TZIS even at constant $\frac{d\tilde\mu_Q}{d\mu_B}$.

\begin{figure}[t]
\centering
\includegraphics[width=0.9\columnwidth]{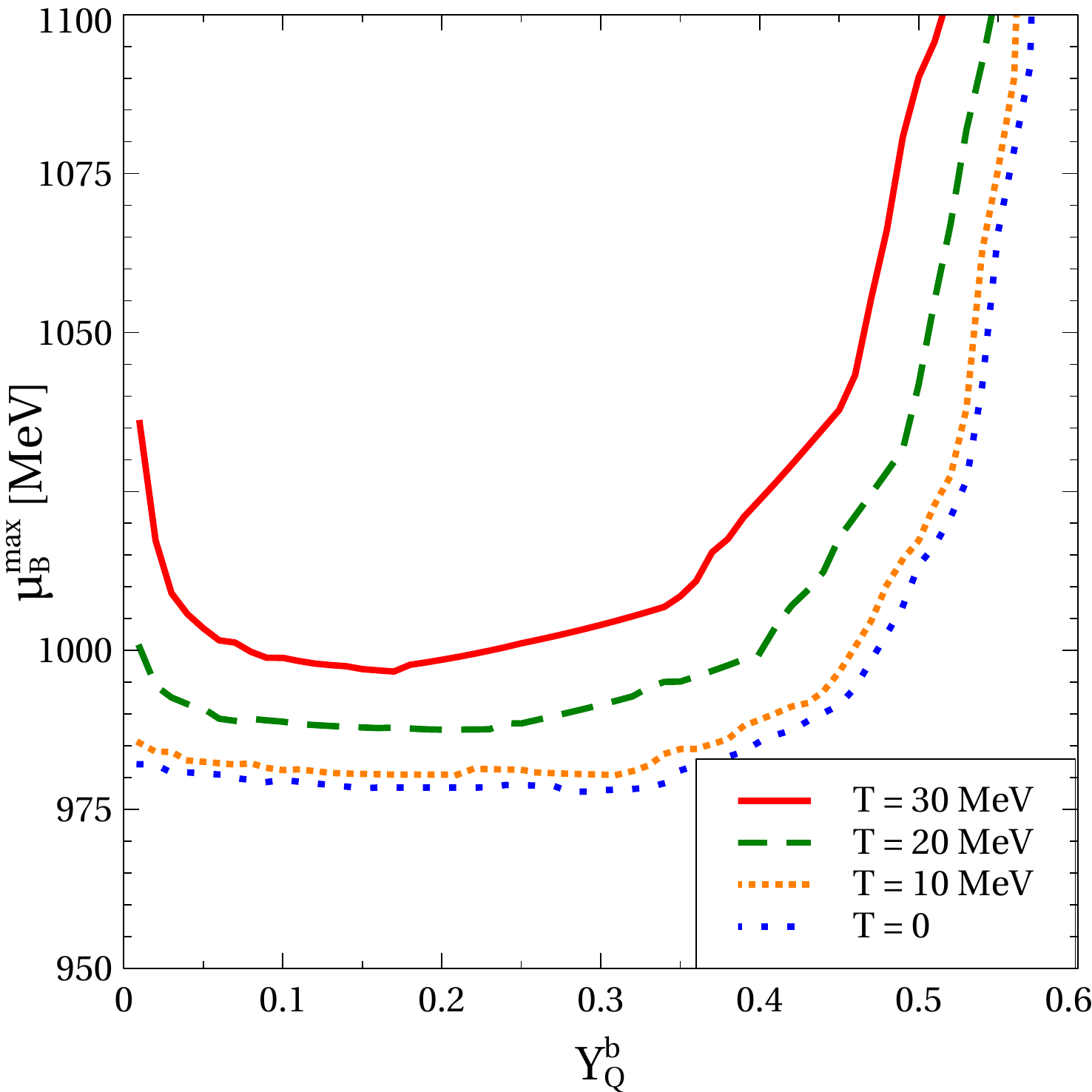}
\caption{Baryon chemical potential of quark-hadron transition $\mu_B^{max}$ found within the Maxwell construction under the condition of $\beta$-equilibrium for several values of temperature $T$.}
\label{fig1}
\end{figure}

\begin{figure}[t]
\centering
\includegraphics[width=0.9\columnwidth]{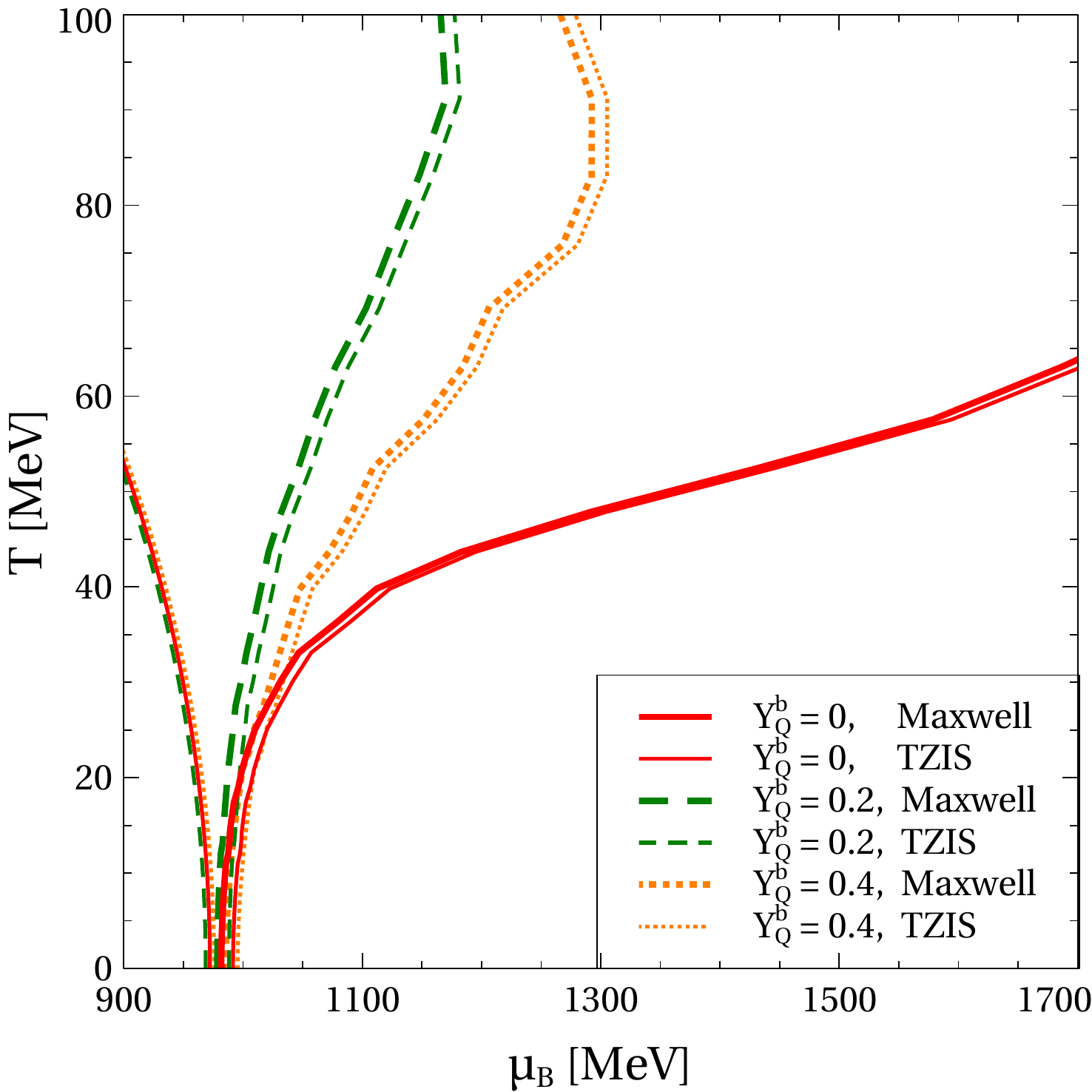}
\caption{Phase diagram of $\beta$-equilibrated quark-hadron matter in the plane of baryon chemical potential $\mu_B$ vs. temperature $T$ at the charge fraction carried by baryons $Y_Q^b=0.01$ (red solid curves), $Y_Q^b=0.2$ (green dashed curves) and $Y_Q^b=0.4$ (orange dotted curves). The curves are obtained by the Maxwell construction (thick curves) and TZIS (thin curves) between EoSs of quark and hadron phases.}
\label{fig2}
\end{figure}

Before going further we would like to show that expansion given by Eqs. (\ref{XXV}) and (\ref{XXVI}) agrees with the thermodynamic identities $n_{B,Q}=\frac{\partial p}{\partial \mu_{B,Q}}$. Total derivative of $\tilde p_{h,q}$ at $T=const$ is
\begin{eqnarray}
\label{XXVII}
\frac{d \tilde p^{h,q}}{d\mu_B}=
\frac{\partial\tilde p^{h,q}}{\partial\mu_B}+
\frac{\partial\tilde p^{h,q}}{\partial\mu_Q}\frac{d\tilde\mu_Q}{d\mu_B}
=\left(1+Y_Q\frac{d\tilde\mu_Q}{d\mu_B}\right)\tilde n_B^{h,q},
\end{eqnarray}   
where on the second step this derivative was explicitly found from Eq. (\ref{XXV}) and rewritten using Eq. (\ref{XXVI}). From this expression we immediately conclude that
\begin{eqnarray}
\label{XXVIII}
\frac{\partial\tilde p^{h,q}}{\partial\mu_B}&=&
\frac{d\tilde p^{h,q}}{d\mu_Q}\biggl|_{\frac{d\tilde\mu_Q}{d\mu_B}=0}
=\tilde n_B^{h,q},\\
\label{XXIX}
\frac{\partial\tilde p^{h,q}}{\partial\mu_Q}&=&
\frac{\frac{d\tilde p^{h,q}}{d\mu_B}-\frac{\partial\tilde p^{h,q}}{\partial\mu_B}}{\frac{d\tilde\mu_Q}{d\mu_B}}=
Y_Q\tilde n_B^{h,q}=\tilde n_Q^{h,q}.
\end{eqnarray}   

The TZIS has six parameters $c_0^h$, $c_0^q$, $c_1^h$, $c_1^q$, $c_2^h$ and  $c_2^q$. By requiring continuity of pressure and density at the mixed phase boundaries we immediately exclude four of the
\begin{eqnarray}
\label{XXX}
c_0^{h,q}&=&p^{h,q}|_{h,q},\\
\label{XXXI}
c_1^{h,q}&=&n^{h,q}_B|_{h,q}.
\end{eqnarray}
Pressure is continuous at $\mu_B=\mu_B^c$, while baryon density can experience a discontinuous jump of the amplitude $\Delta n_B$, i.e.
\begin{eqnarray}
\label{XXXII}
\tilde p^h|_c&=&\tilde p^q|_c,\\
\label{XXXIII}
\tilde n_B^h|_c&=&\tilde n_B^q|_c-\Delta n_B.
\end{eqnarray}
The amplitude $\Delta n_B$ is finite at the first order phase transition and vanishes at the second order one or for a cross-over. 
Therefore, by specifying $\Delta n_B$ as a function of temperature $T$ we can model the strongly interacting matter phase diagram and its CEP(s) phenomenologically. 
In addition to the high temperature CEP at $T\sim 100$ MeV 
\footnote{According to lattice QCD simulations, the high-temperature CEP, if it exists at all, has to occur at $T_{cep1}<132^{+3}_{-6}$ MeV \cite{HotQCD:2019xnw}.},
it is reported that the interplay between scalar diquark and chiral condensates leads to the appearance of another CEP at low temperatures \cite{Hatsuda:2006ps}. 
Within this scenario, $\Delta n_B>0$ at $T_{cep1}>T>T_{cep2}$ and $\Delta n_B=0$ elsewhere. 
Introducing $t_1=T/T_{\rm cep1}-1$ and $t_2=T/T_{\rm cep2}-1$, we parameterize the density jump as
\begin{eqnarray}
\label{XXXIV}
\Delta n_B=n^*|t_1|^{\beta_1}|t_2|^{\beta_2}\theta(t_1)\theta(t_2),
\end{eqnarray}
where the constant $n^*$ controls its amplitude and $\beta_1$, $\beta_2$ are critical exponents. 
These exponents are fixed to their value in models of the 3D Ising universality class \cite{Campostrini:2002cf}, i.e. $\beta_1=\beta_2=0.3265$.
This value also falls into the range $\beta=0.32-0.35$ of simple liquids \cite{Huang_1987}. 
We consider the cases of $n^*=0$ and $n^*=0.15~{\rm fm}^{-3}$, while the critical temperatures are assigned the values $T_{\rm cep1}=90$ MeV and $T_{\rm cep2}=15$ MeV. 

\begin{figure}[t]
\centering
\includegraphics[width=0.9\columnwidth]{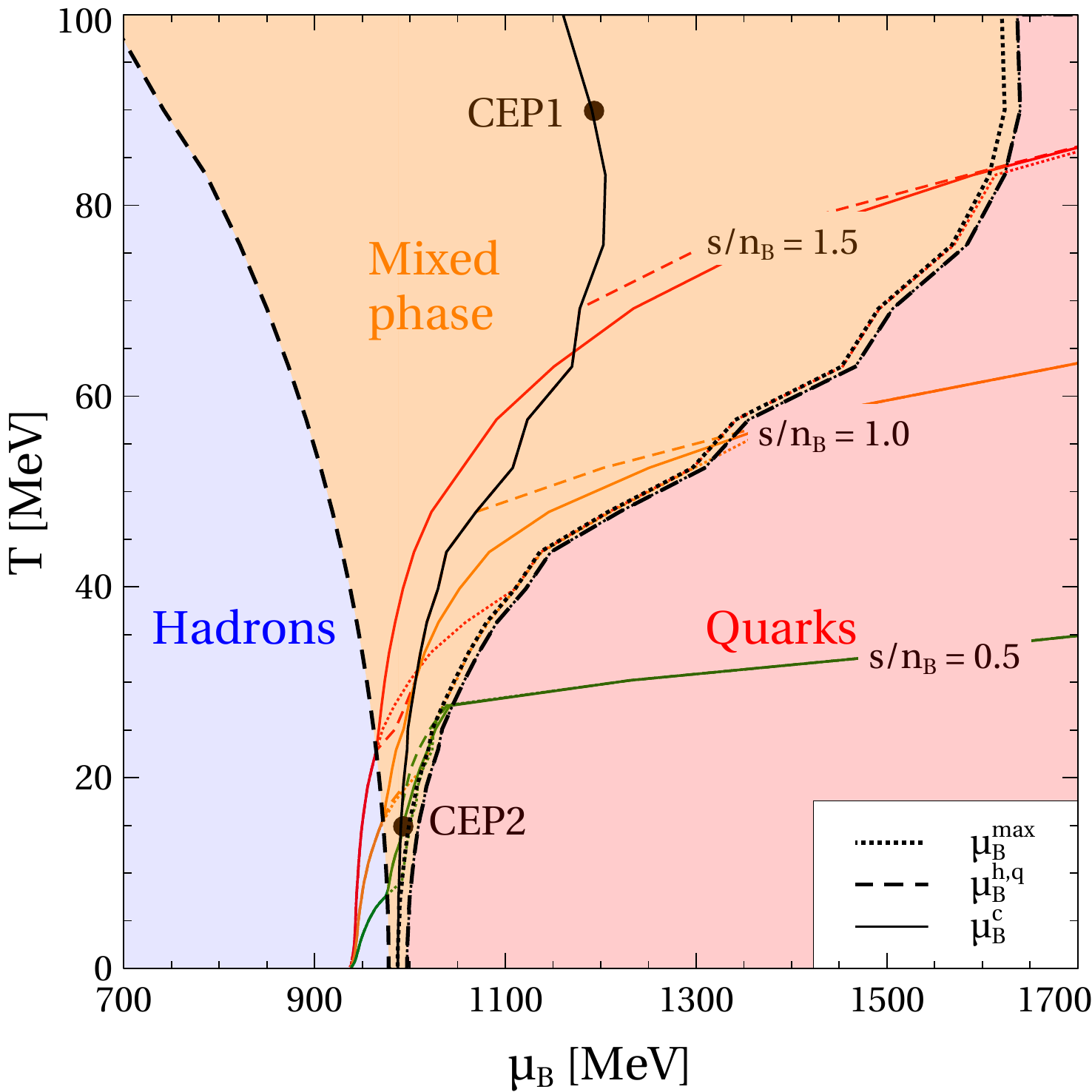}
\caption{Phase diagram of $\beta$-equilibrated electrically neutral quark-hadron matter in the plane of baryon chemical potential $\mu_B$ vs. temperature $T$. The black dotted, dashed and solid curves correspond to the phase boundary of the Maxwell construction, the phase boundaries of the TZIS and the merging chemical potential of the TZIS, respectively. 
The color mapping of hadron, quark and mixed phases corresponds to TZIS. The filled black circles show the CEPs. 
The colored curves represent adiabates with values of $s/n_{B}$ calculated within the Maxwell construction (dotted curves), the TZIS with $n^*=0.15~fm^{-3}$ (dashed curves) and the TZIS with $n^*=0$ (solid curves).}
\label{fig3}
\end{figure}

\begin{figure}[t]
\centering
\includegraphics[width=0.9\columnwidth]{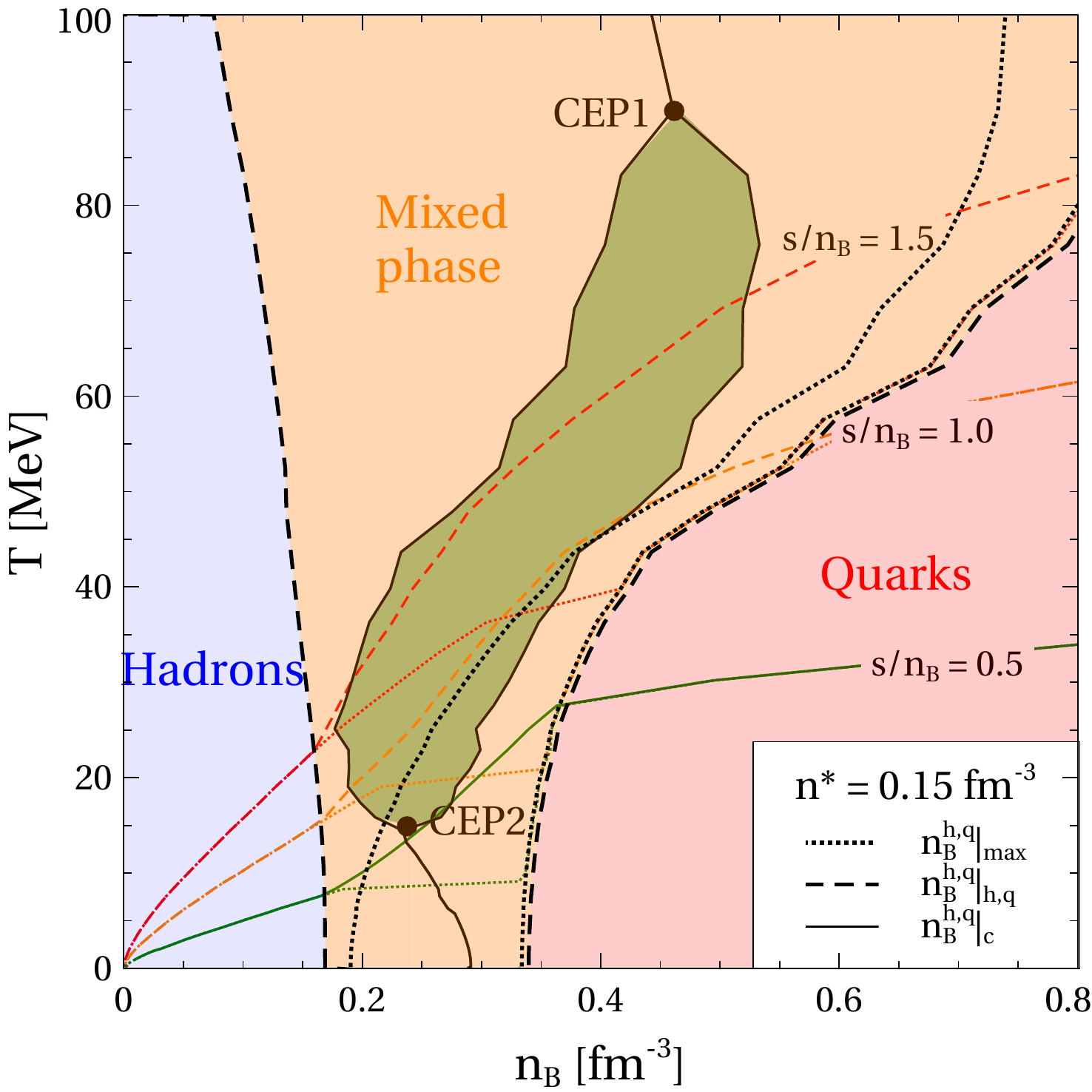}
\includegraphics[width=0.9\columnwidth]{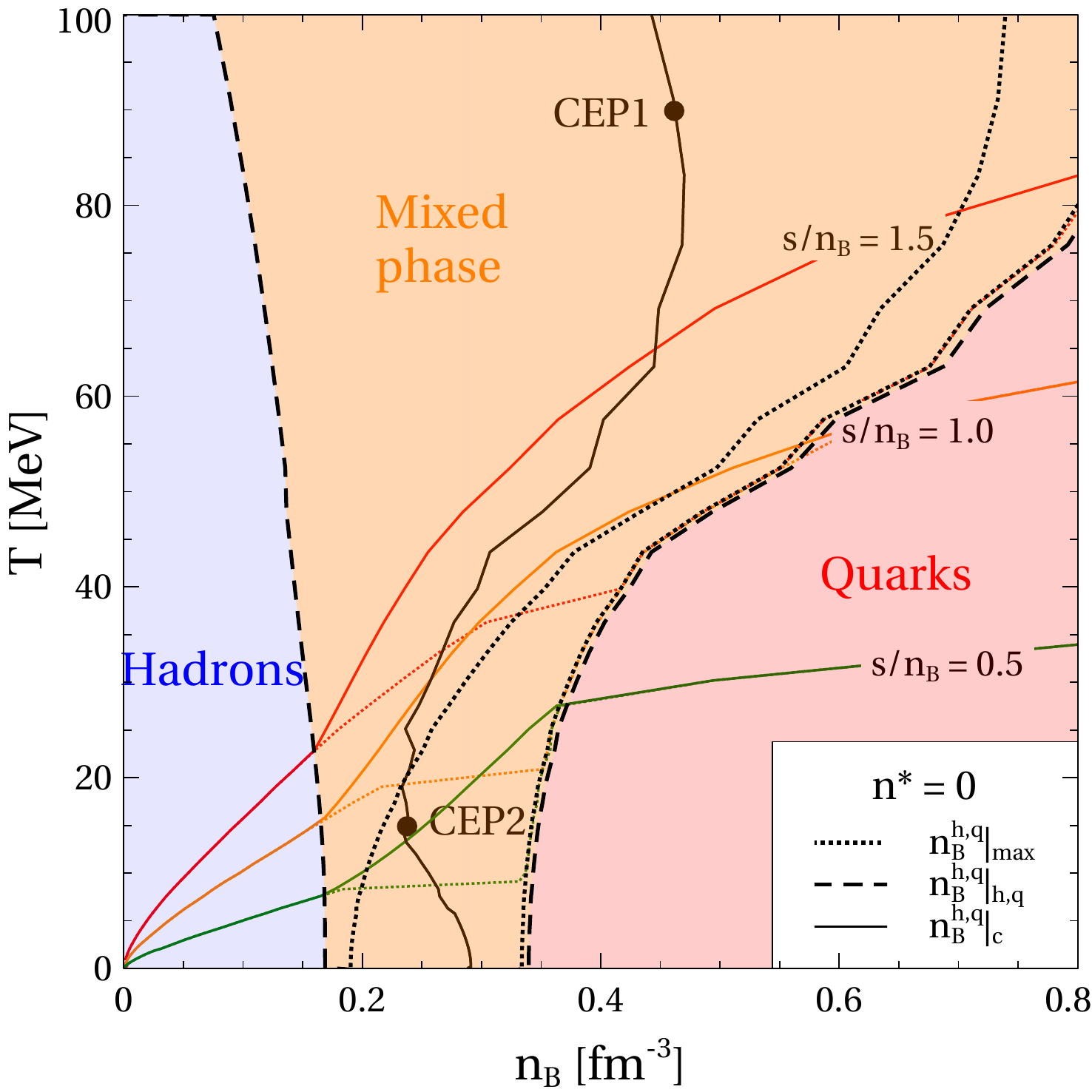}
\caption{The same as on Fig. \ref{fig3} but in the plane of baryon density $n_B$ vs. temperature $T$ at $n^*=0.15~fm^{-3}$ (upper panel) and $n^*=0$ (lower panel). Green shading on the upper panel demonstrates the region where the baryonic density experiences a discontinuous jump within the TZIS. Low and high temperature CEPs on the lower panel are shown in order to guide the eye.}
\label{fig4}
\end{figure}

The quark-hadron mixed phase can be characterized by the volume fraction of quark matter $\lambda^{h,q}\in[0,1]$, which should be defined for the hadron (superscript index ``$h$'') and quark (superscript index ``$q$'') branches of the TZIS. 
Within this notation the volume fraction of hadronic matter is $1-\lambda^{h,q}$. 
The volume fraction $\lambda^{h,q}$ is related to the baryonic charge density of mixed phase as $\tilde n_B^{h,q}=\lambda^{h,q}\tilde n_B^h|_h+(1-\lambda^{h,q})\tilde n_B^q|_q$. 
This relation allows us to find
\begin{eqnarray}
\label{XXXV}
\lambda^{h,q}=\frac{\tilde n_B^{h,q}-\tilde n_B^h|_h}{\tilde n_B^q|_q-\tilde n_B^h|_h},
\end{eqnarray}
which provides a direct access to the relevant thermodynamic quantities of the mixed phase. For example, the mixed phase entropy density reads 
\begin{eqnarray}
\label{XXXVI}
\tilde s^{h,q}=\lambda^{h,q}\tilde s^h|_h+(1-\lambda^{h,q})\tilde s^q|_q.
\end{eqnarray}
The energy density needed as an input for solving the TOV equation can be found using the thermodynamic identity
\begin{eqnarray}
\tilde\epsilon^{h,q}&=&T\tilde s^{h,q}+\mu_B\tilde n_B^{h,q}+\tilde \mu_Q\tilde n_Q^{h,q}-\tilde p^{h,q}\nonumber\\
\label{XXXVII}
&=&
T\tilde s^{h,q}+(\mu_B+Y_Q\tilde \mu_Q)\tilde n_B^{h,q}-\tilde p^{h,q},
\end{eqnarray}
where in the second step the electric charge density was expressed through the baryonic charge density and the electric charge fraction.

\begin{figure}[!]
\centering
\includegraphics[width=0.8\columnwidth]{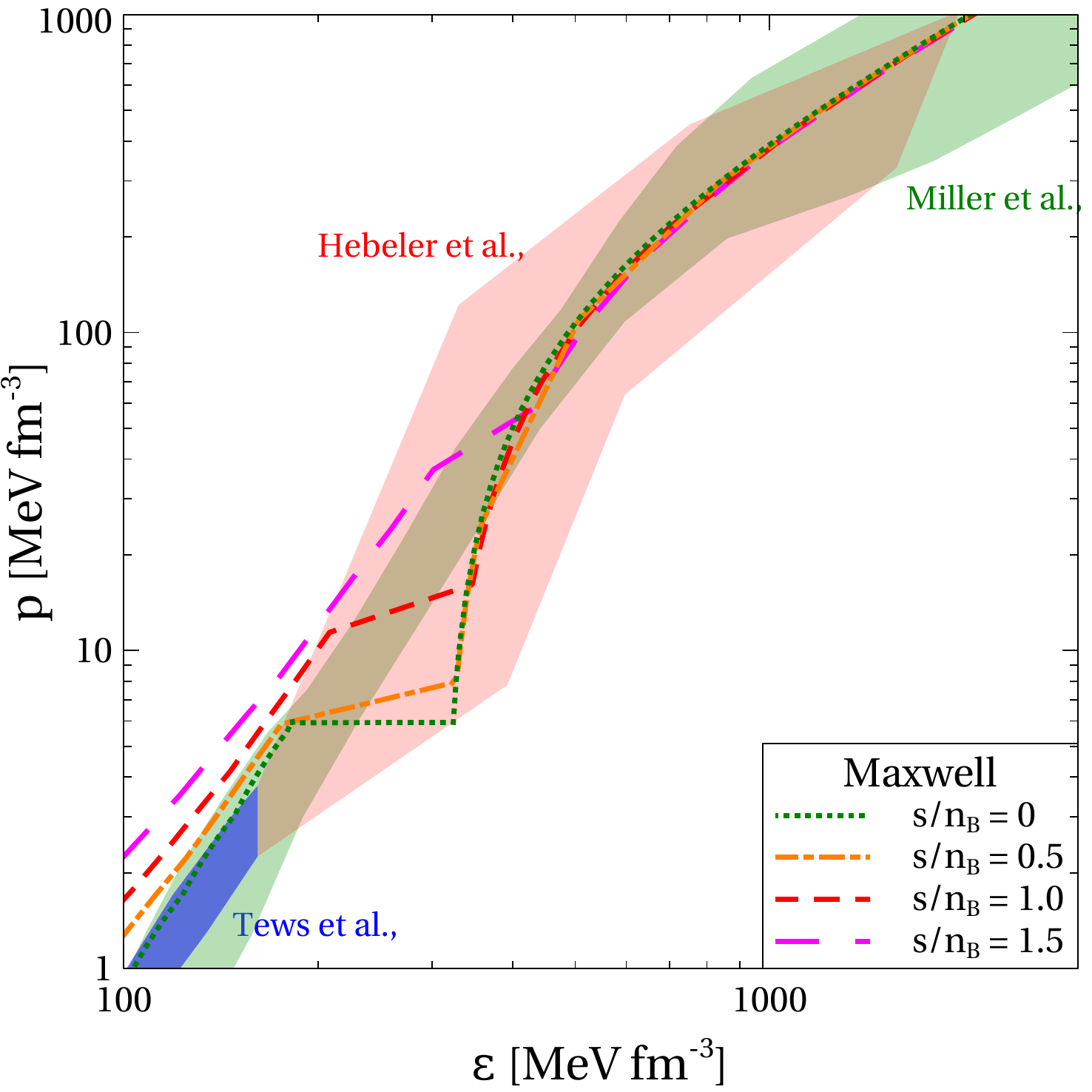}
\includegraphics[width=0.8\columnwidth]{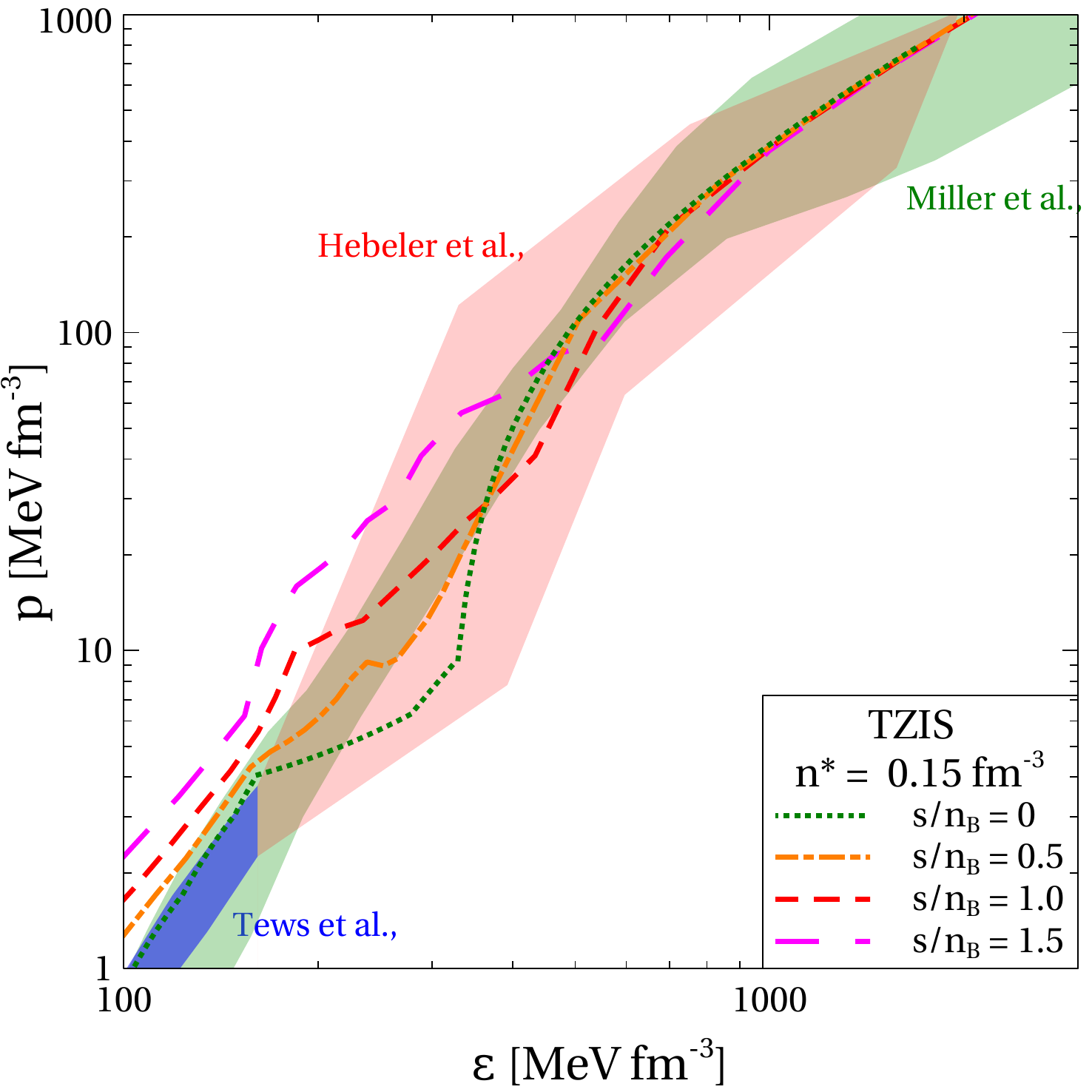}
\includegraphics[width=0.8\columnwidth]{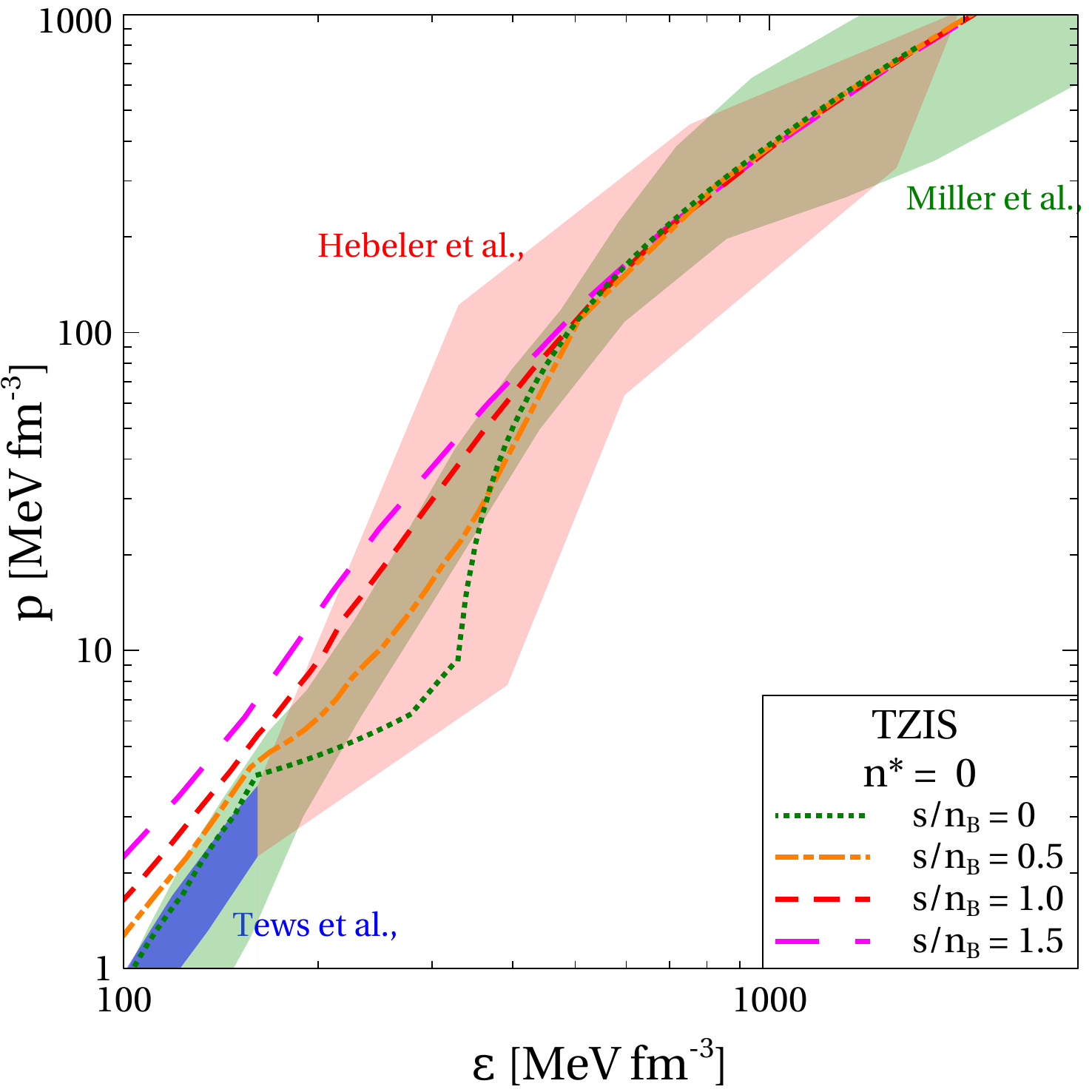}
\caption{EoS of electrically neutral $\beta$-equilibrated quark-hadron matter in the plane of energy density $\varepsilon$ and pressure $p$ along the trajectories of constant entropy per baryon $s/n_B$ found within the Maxwell construction (upper panel) and the TZIS with $n^*=0.15~fm^{-3}$ (middle panel) and $n^*=0$ (lower panel). The nuclear matter constraints represented by the shaded areas are discussed in the text.}
\label{fig5}
\end{figure}

\subsection{Phase diagram}

We first analyze the effect of the electric charge fraction carried by baryons $Y_{Q}^b$ on the shape of the phase diagram of strongly interacting matter in $\beta$-equilibrium. 
For simplicity only electrons are taken into consideration. 
In $\beta$-equilibrium, their chemical potential is $\mu_e=-\mu_Q$. 
More generally, the chemical potential of a particle with baryonic charge $B$ and electric charge $Q$ is $\mu=B\mu_B+Q\mu_Q$. 
Fig.~\ref{fig1} shows the behavior of the chemical potential of the quark-hadron transition under the Maxwell construction $\mu_{B}^{\rm max}$ as a function of $Y_{Q}^b$. 
It is seen that $\mu_{B}^{\rm max}$ is not monotonous but has a minimum at all values of temperature $T$. 
The same qualitative conclusion holds for $\mu_B^h$ and $\mu_B^q$, which 
are not shown in Fig.~\ref{fig1} for the sake of clarity. 
In other words, a certain value of $Y_{Q}^b$ leads to the smallest density of the quark matter onset. 

Fig.~\ref{fig2} shows the phase diagram of $\beta$-equilibrated quark-hadron matter found by the Maxwell construction and by the TZIS. 
The phase coexistence curve found within the Maxwell construction has a characteristic shape with $\mu_B^{max}$ shifted toward small values at low temperatures. This is due to the lowering of the onset density of quark matter deconfinement caused by diquark pairing, which is most pronounced at small $T$. 
A similar behavior is observed for the TZIS mixed phase boundary from the quark side $\mu_B^q$, which is strongly correlated with $\mu_B^{\rm max}$. For the TZIS merging chemical potential $\mu_B^c$ this effect is also present but appears to be less pronounced due to the normal behavior of the TZIS mixed phase boundary from the hadron side $\mu_B^h$.  
At the same time we would like to stress that the temperature derivatives of $\mu_B^h$, $\mu_B^c$ and $\mu_B^q$ vanish at $T=0$. 
The hadron boundary $\mu_B^h$ found with Eq.~(\ref{XIV}) weakly depends on $Y_Q^b$, while the quark one $\mu_B^q$ is strongly sensitive to the value of the electric charge fraction carried by baryons. 
It is important to note, that for the analyzed range of temperatures the width of the mixed phase region in the TZIS grows with $T$ at any $Y_Q^b$. 

\begin{figure}[!]
\centering
\includegraphics[width=0.8\columnwidth]{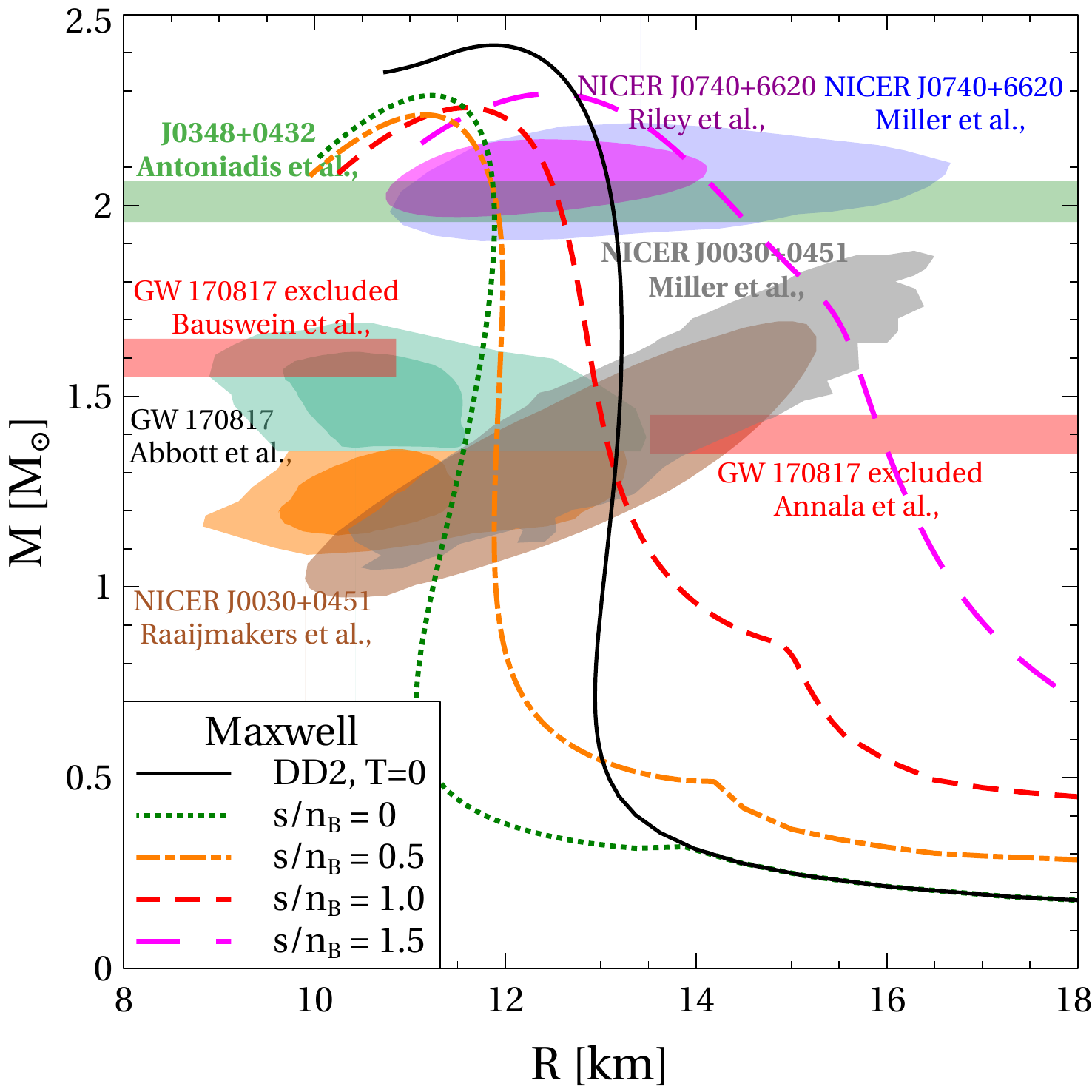}
\includegraphics[width=0.8\columnwidth]{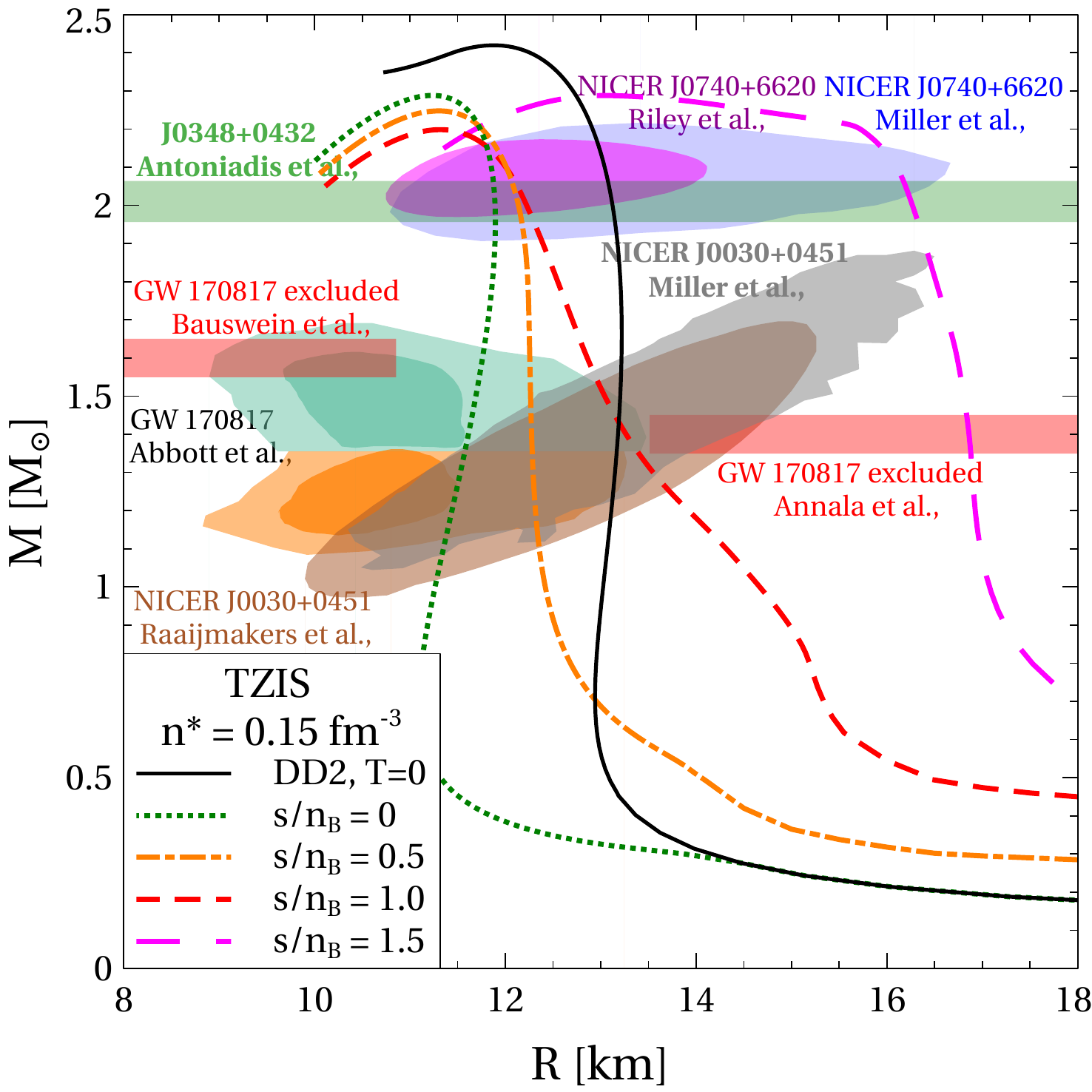}
\includegraphics[width=0.8\columnwidth]{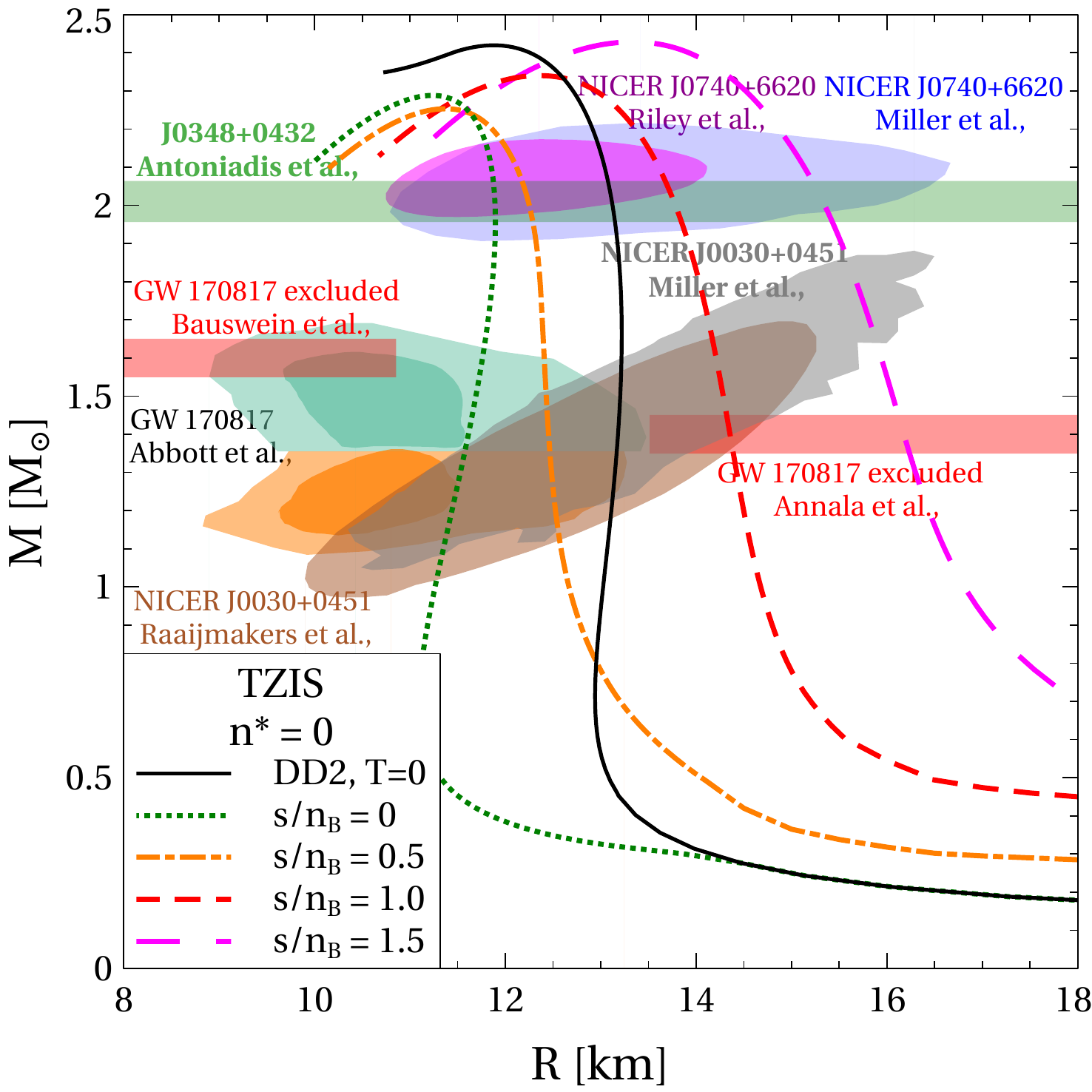}
\caption{Mass-radius relation of hybrid neutron stars with the isentropic quark-hadron EoS presented in Fig.~\ref{fig5} compared to the mass-radius relation obtained with the DD2 EoS of cold hadron matter. The astrophysical constraints depicted by the colored bands and shaded areas are discussed in the text.}
\label{fig6}
\end{figure}
 
Modelling (proto)neutron stars requires an additional condition on the EoS of stellar matter, i.e. the electric charge neutrality provided by a proper amount of electrons. 
This corresponds to solving $Y_Q=0$ with respect to $\mu_Q$. 
Fig.~\ref{fig3} shows the corresponding phase diagram in the plane of baryonic chemical potential and temperature. 
Fig.~\ref{fig3} also demonstrates the trajectories of constant entropy per baryon $s/n_B$ of $\beta$-equilibrated electrically neutral quark-hadron matter. 
A remarkable and general feature of the shown trajectories is temperature increase caused by transition from hadron phase to the quark one observed at any nonzero $s/n_B$ within the Maxwell construction and the TZIS. 
This is a direct consequence of the reducing the number of microstates due to diquark pairing as compared to the case of unpaired quark matter. 
Indeed, each of the red and green quarks participating in the pairing can exist in two spin, two color and two flavor states, while a spin-color singlet diquark has just one available state. 
Thus, a reduction of the number of available microstates requires an increase of temperature in order to conserve the entropy. 
Fig.~\ref{fig4} shows the same phase diagram in the plane of baryonic density and temperature. 
The shape of the mixed phase boundaries is qualitatively similar to the one in the $\mu_B-T$ plane and is not affected by the value of $n^*$. 
It is seen from the upper panel that at non-zero values of this parameter the temperature range between $T_{\rm cep1}$ and $T_{\rm cep2}$ includes the domain where $n_B$ experiences a discontinuous jump. 
It is also interesting to note, that at any $n^*$ the isentropic trajectories of the mixed phase found within the TZIS are located above the ones obtained with the Maxwell construction. 
This means that a reduction of the available number of micriostates due to the transition form hadron matter to color-superconducting quark matter in the TZIS is more pronounced than within the Maxwell construction because of the larger volume fraction of quark matter. 

\section{Protoneutron stars with quark cores}
\label{sec4}

The quark-hadron matter in the interiors of the protoneutron stars that are created during supernovae explosions evolves along the trajectories which are approximately isentropic \cite{Fischer:2017lag}. 
Therefore for modelling these astrophysical objects we consider hybrid EoS of $\beta$-equilibrated electrically neutral matter calculated under the condition of constant ratio $s/n_B$. Fig. \ref{fig5} shows the corresponding pressure as a function of the energy density. The first conclusion valid for both TZIS and Maxwell construction is that growth of the entropy per baryon and, consequently, temperature leads to increase of pressure at a given value of energy density. This stiffening of the quark-hadron EoS is the most pronounced at low densities, while at high $\varepsilon$ effects of $s/n_B$ and $T$ are relatively weak. Switching from hadron matter to the quark one leads to softening of the EoS in the mixed phase region at any $s/n_B$. The stronger is phase transition, the more pronounced is this effect being the most spectacular in the case of the Maxwell construction, while TZIS with vanishing $n^*$ diminishes it. At $s/n_B=0$ our hybrid EoS obtained within the Maxwell construction and TZIS with vanishing and finite $n^*$ agrees with the low density calculations of the chiral EFT approach \cite{Kruger:2013kua} and constraints from the multipolytrope analysis of the PSR J1614+2230 \cite{Hebeler:2013nza} and PSR J0740+6620 \cite{Miller:2021qha} observational data. As expected, at finite $s/n_B$ this agreement is spoiled in the low density region.

We apply the developed hybrid EoSs as an input to the problem of relativistic hydrostatic equilibrium, i.e. to solving the TOV equation giving a mass-radius relation of protoneutron stars. This relation is shown on Fig. \ref{fig6}. We compare it to the constraint on the lower limit of the TOV maximum mass given by the mass $2.01^{+0.04}_{+0.04}~\rm M_\odot$ measured in a binary system of the pulsar PSR J0348+0432 and its white dwarf companion \cite{Antoniadis:2013pzd}, to results of the Bayesian analysis of the observational data from PSR J0740+6620 \cite{Riley:2021pdl,Miller:2021qha} and PSR J0030+0451 \cite{Riley:2019yda,Raaijmakers:2019qny}, analysis of the gravitational wave signal produced by the merger GW170817 \cite{LIGOScientific:2018cki} as well as limitations on the stellar radius at $1.6~\rm M_\odot$ from below by $R_{1.6}\ge 10.68$ km \cite{Bauswein:2017vtn} and at $1.4~\rm M_\odot$ from above by $R_{1.4}\le 13.6$ km \cite{Annala:2017llu}. 
The zero-entropy hybrid EoS obtained with the Maxwell construction and with the TZIS fits all these constraints and is used as a benchmark. 
Increasing $s/n_B$ shifts the mass-radius diagram towards large radii, while leaving the maximum mass almost unchanged. This is due to the fact that entropy effects are most pronounced in the low density regime, while being small at high densities. 
Nevertheless, all the constraints mentioned above are fulfilled at $s/n_B=0.5$. At $s/n_B=1.0$ this is the case only within the Maxwell construction of the quark-hadron transition, while the TZIS provides a marginal agreement only for $n^*=0.15~fm^{-3}$. 
A further increase of the entropy per baryon leads to large values for the protoneutron star radii, e.g., $R_{1.4}\simeq 16-17$ km for $s/n_B=1.5$. 

\section{Conclusions}
\label{sec5}

We have developed the generalization to finite temperatures of a recently proposed two-zone interpolation scheme that matches the domains of pure hadronic and quark matter phases and thus allows to study the phase diagram of strongly interacting matter. 
The extension of the approach to the case of a finite fraction of electric charge is a novel element of the presented work. 
We investigated how this parameter modifies the shape of the phase boundary. We also considered two scenarios of the quark-hadron transition, namely a continuous and a discontinuous change of the baryon density at the transition, corresponding to a cross-over and a strong first order phase transition, respectively. Within the second scenario the phase transition curve is terminated at the low and high temperature critical endpoints. 

An important aspect of the study is incorporation of color superconductivity based on the approach of a confining density functional for quark matter. The formation of a color superconducting state of quark matter is responsible for a characteristic shape of the mixed phase boundary in the case of the Maxwell construction and two-zone interpolation scheme. 
Another important effect of color superconductivity which is absent in the case of normal quark matter, is the growth of the temperature at the transition from the hadronic phase to the quark matter phase. This effect drives the trajectories of evolution of protoneutron stars produced in the supernova explosions toward the regions of the phase diagram that are accessed in the NS mergers and in the laboratory experiments with collisions of relativistic heavy ions.

Finally, we analysed the effects of entropy on the mass-radius relation of protoneutron stars with quark-hadron transition within the Maxwell construction and the two-zone interpolation scheme with a first order phase transition and cross-over. A finite entropy per baryon strongly modifies stellar radii while leaving the maximum mass almost unchanged. 

Despite the fact that the present study is mostly focused on the temperature-density region typical for astrophysical applications, the proposed approach can be applied to the entire phase diagram of strongly interacting matter.

\subsection*{Acknowledgments}

The authors acknowledge support from the Polish National Science Centre (NCN) under grant number\\ 2019/33/B/ST9/03059. 
This work is part of a project that has received funding from the European Union’s Horizon 2020 research and innovation program under the grant agreement STRONG – 2020 - No 824093.
We are grateful to the COST Action CA16214 "PHAROS" for supporting our networking activities. We also express our acknowledgements to Tobias Fischer for fruitful discussions, Alexander Ayriyan for indispensable help with numerical calculations and Mahboubeh Shahrbaf for valuable assistance in preparing the paper.

\bibliography{epja}

\end{document}